\documentclass[]{interact}

\usepackage{color}
\usepackage{graphicx}
\usepackage{mathtools}
\usepackage{siunitx}
\usepackage{multirow}
\usepackage{orcidlink}
\usepackage{hyperref}
\hypersetup{colorlinks,citecolor=blue,linkcolor=blue,urlcolor=blue}
\graphicspath{{./fig/}}
\DeclareMathOperator{\arctanh}{arctanh}

\usepackage{epstopdf}
\usepackage{subfigure}

\usepackage{natbib}
\bibpunct[, ]{(}{)}{,}{a}{}{,}
\renewcommand\bibfont{\fontsize{10}{12}\selectfont}

\theoremstyle{plain}

\theoremstyle{definition}

\theoremstyle{remark}

\newcommand{\hourglass}{Hourglass}

\begin{document}

\articletype{ARTICLE TEMPLATE}

\title{Hourglass: an unpublished map projection by John P. Snyder}

\author{
\name{J. Jimenez Shaw\textsuperscript{a}%
      \thanks{CONTACT J. Jimenez Shaw. Email: javier.shaw@pix4d.com}%
      \orcidlink{0000-0002-7227-9173} and
      J.L.G. Pallero\textsuperscript{b}%
      \thanks{CONTACT J.L.G. Pallero. Email: jlg.pallero@upm.es}%
      \orcidlink{0000-0001-8158-106X}}
\affil{\textsuperscript{a}Pix4D GmbH, Berlin, Germany;
       \textsuperscript{b}ETSI en Topograf\'ia, Geodesia y Cartograf\'ia.
                          Universidad Polit\'ecnica de Madrid, Madrid, Spain}
}

\maketitle

\begin{abstract}
Hourglass is an equal-area pseudocylindrical map projection developed by John P.
Snyder in mid 1940s. It was never published in a detailed way by its author, and
only a couple of references exist in literature since $1991$, both of them 
including a picture but without any mathematical description. In this work the 
equations for the ellipsoid and for the sphere are derived, both for direct and 
inverse problems, together with a generalization that allow meridians not 
restricted to straight lines as in the original Snyder's version. Although not
useful for world maps, the projection can be employed for mapping areas around 
their standard parallels.
\end{abstract}


\section{Introduction}
\label{sec:Intro}

In \textit{How to Lie with Maps}, by Mark Monmonier, a map projection called 
\textit{\hourglass}, attributed to John P. Snyder, is mentioned as an example
to make the point that area fidelity does not necessarily confer shape
fidelity \citep[p.98]{monmonier1991}. In other words, an equal-area projection
is not always a \textit{good} projection. In his book of $2004$ 
\textit{Rhumb Lines and Map Wars}, the same author shows again a picture of the 
\hourglass{} projection together with an excerpt from a letter from Snyder to 
Arthur Robinson which reads \citep[pp. 162--163]{monmonier2004}:
\begin{quotation}
Enclosed is an equal-area pseudocylindrical-type projection I devised around
$1946$, and drew by hand then. I felt it should remain buried in my notes...
\end{quotation}
The projection is also mentioned in \citet[p. 308, note 107]{snyder1997}, but 
referencing \cite{monmonier1991} and without any details qualitative nor 
quantitative.

\begin{figure*}[htb]
\centering
\includegraphics[width=0.85\textwidth]{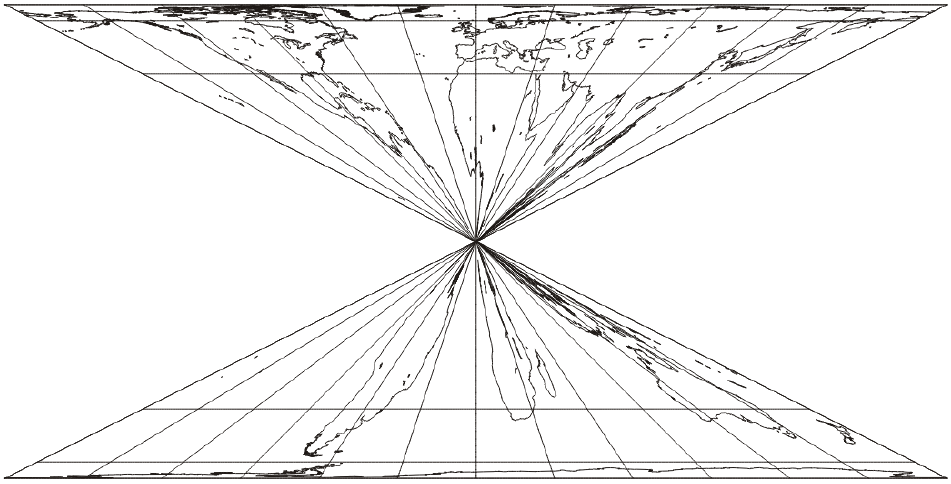}
\caption{Hourglass projection as drawn in \citet[p. 98]{monmonier1991} and
         \citet[p. 163]{monmonier2004}.}
\label{fig:figure1}
\end{figure*}

Figure~\ref{fig:figure1} shows the appearance of the world in the Snyder's
equal-area \hourglass{} map projection as it is drawn in \cite{monmonier1991} and
\cite{monmonier2004}. The equator is a single point (the origin of the map) and
the poles are straight lines. Each hemisphere is an isosceles triangle, being
their non-equal angles vertically opposite. Parallels are drawn as
straight lines parallel to $X$ axis, while meridians are straight lines that
converge all at the central point at the equator (the $Y$ axis is the origin
meridian $\lambda_0$). Along a certain projected parallels $\pm\varphi_0$, distances
are true.

It is clear that this projection does not have almost any practical utility, at
least from a professional point of view to make a functional world map, although
it could be useful for latitudes around the standard parallel. However, we want
to point out other aspects, such as, as Snyder did, show how we should be
careful judging isolately this or that property from a projection (Snyder had a
clear target at that time: criticism the Peters projection). As it is stated in
\citet[p. 161]{monmonier2004},
\begin{quotation}
... a map's effectiveness as an ideological statement does not make it a
reliable device for representing area, shape, or relative importance. Favor one
role or attribute, and you're likely to slight others.
\end{quotation}

Pseudocylindrical projections are characterized by straight horizontal lines for
parallels and (usually) equally-spaced curved meridians \citep{snyder1977}.
Although \hourglass{} meridians are not curved lines, the projection can be
classified as pseudocylindrical in the same way that the Eckert II or the
Collignon projections are also. Hourglass would belong then to the rectilinear
pseudocylindricals subcategory stated in \cite{snyder1977}. However, as it was
stated previously, no mathematical description for the \hourglass{} projection is
shown in the cited references; none of them include any formulas that allow us
to replicate the presented maps, and this is the reason for the 
\textit{unpublished} term in the title of this paper. In this work we will 
develop the formulation for the ellipsoid and the sphere not only for the 
original Snyder's \hourglass{} projection, but also for a generalization of it 
that allows  non straight meridians.

\section{General equations and aspect ratio}
\label{sec:AR}

Consider a generic map projection defined as
\begin{equation}
\label{ec:xfyh}
\begin{dcases}
x=f(\varphi,\lambda),\\
y=h(\varphi,\lambda),
\end{dcases}
\end{equation}
where $\varphi$ is geodesic latitude and $\lambda$ geodesic longitude. Consider
the origin of the plane coordinates, according to Figure~\ref{fig:figure1}, at
the equator single point, being the $Y$ axis the North-South direction and the
$X$ axis the East-West direction. As it can be seen in
Figure~\ref{fig:figure1}, projected parallels are straight lines parallel to the
$X$ axis, while projected meridians are straight lines convergent in the origin.
In these circumstances, we can state that $y$ coordinate of a point is only
latitude dependent, while $x$ coordinate is both latitude and longitude
dependent. So, Equation~(\ref{ec:xfyh}) can be written as
\begin{equation}
\label{ec:xfyhFin}
\begin{dcases}
x=f(\varphi,\lambda),\\
y=h(\varphi),
\end{dcases}
\end{equation}
belonging \hourglass{} then to Tobler's category C projections \citep{tobler1962}.

\begin{figure}[htb]
\centering
\includegraphics[width=0.35\textwidth]{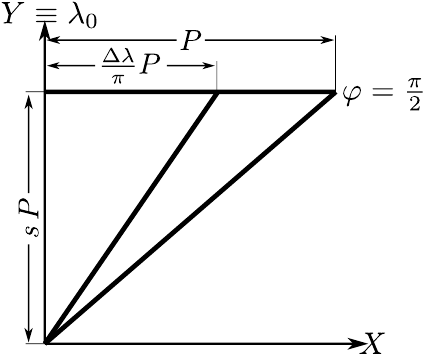}
\caption{Projection of the origin meridian $\lambda_0$ and the pole.}
\label{fig:figure2}
\end{figure}

We introduce now a parameter $s>0$ in order to control the relationship between
the horizontal and vertical dimensions of the map. Figure~\ref{fig:figure2}
presents the projection of the origin meridian $\lambda_0$ and the North pole, 
which is a horizontal straight line of length $P$ for $\Delta\lambda=\pi$. If we impose the $y$ coordinate 
of the pole as $y=s\,P$, then an aspect ratio of $s:1$ means that the height of the
whole map is $s$ times its whole width. For any point with longitude
$\Delta\lambda=\lambda-\lambda_0$ its $x$ coordinate at the line representing the
pole is $x=\Delta\lambda\,P/\pi$.

Meridians are straight lines, so
\begin{equation}
y=c\,x.
\end{equation}
For any point of longitude $\Delta\lambda$ the slope of the line representing
its meridian is then
\begin{equation}
\label{ec:DefCOrig}
c=\frac{y}{x}=\frac{s\,P}{\Delta\lambda\,P/\pi}=\frac{s\,\pi}{\Delta\lambda}.
\end{equation}
Then, the $x$ and $y$ projected coordinates are related as
\begin{equation}
\label{ec:xGeneral}
x=\frac{\Delta\lambda}{s\,\pi}|y|,
\end{equation}
where the absolute value in $y$ guarantees correct results for northern and
southern hemispheres whenever $\Delta\lambda$ keeps the correct sign (positive
eastwards $\lambda_0$ and negative westwards $\lambda_0$).

\section{Equal-area condition}

For an ellipsoid of semimajor axis $a$ and first eccentricity $e$, an equal-area
map projection must fulfill for any point that \citep[p. 28]{snyder1987}
\begin{equation}
\label{ec:CondEquiv}
\frac{\partial y}{\partial\varphi}\frac{\partial x}{\partial\lambda}
-\frac{\partial x}{\partial\varphi}\frac{\partial y}{\partial\lambda}=\rho\,r=\rho\,\nu\cos\varphi
=\frac{a^2(1-e^2)}{(1-e^2\sin^2\varphi)^2}\cos\varphi,
\end{equation}
where $\rho$ is the radius of curvature of the meridian ellipse, $r$ is the
radius of the parallel, and $\nu$ is the radius of the first vertical, all of
them at latitude $\varphi$.

According to \cite{snyder1977},
\begin{quotation}
If $y$ is any function of latitude\footnote{We adopt in this excerpt the same 
notation for latitude as in the original reference.} $\phi$, multiplied by the radius 
$R$ of a globe of equal surface area, or
\begin{equation}
\label{ec:Sny1}
y=Rf(\phi),
\end{equation}
then $x$ and $y$ are the rectangular coordinates of an equal-area
pseudocylindrical map projection, where
\begin{equation}
\label{ec:Sny2}
x=\frac{R^2\lambda\cos\phi}{dy/d\phi},
\end{equation}
with $\lambda$ the longitude difference (east is positive) between the given
meridian and the central meridian. All normal pseudocylindrical equal-area
projections must satisfy equations~(\ref{ec:Sny1}) and~(\ref{ec:Sny2}).
\end{quotation}

Equation~(\ref{ec:Sny2}) can be deduced from~(\ref{ec:CondEquiv})
and~(\ref{ec:Sny1}). For a general sphere of radius $R$,
$\rho\,\nu=R^2\cos\varphi$, and taking into account~(\ref{ec:Sny1}) we deduce
that $\partial y/\partial\lambda=0$, so, applying the equal-area condition,
\begin{equation}
\frac{\partial y}{\partial\varphi}\frac{\partial x}{\partial\lambda}
=R^2\cos\varphi\longrightarrow
dx=\frac{R^2\cos\varphi}{\partial y/\partial\varphi}d\lambda,
\end{equation}
which, integrated, arrives to Equation~(\ref{ec:Sny2}). We will show that
\hourglass{} equations have the forms~(\ref{ec:Sny1}) and~(\ref{ec:Sny2}) when
applied to the sphere.

On the other hand, there can be no conformal pseudocylindrical projections
since the curved meridians cannot satisfy the conformal requeriment of
intersecting all straight parallels at right angles \citep{snyder1977}.

\section{Formulation for the ellipsoid}

\subsection{Forward problem}

As it was stated, \hourglass{} projection is equal-area. As the projected parallels
are straight lines parallel to the $X$ axis $y$ coordinate is independent of
$\lambda$, as it was seen in Equation~(\ref{ec:xfyhFin}), so we can write
\begin{equation}
\label{ec:dydlon}
\frac{\partial y}{\partial\lambda}=0.
\end{equation}
On the other hand, from Equation~(\ref{ec:xGeneral}) for the $x$ coordinate
we can write
\begin{equation}
\label{ec:dxdlon}
\frac{\partial x}{\partial\lambda}=\frac{y}{s\,\pi}.
\end{equation}

Then, putting together~(\ref{ec:CondEquiv}), (\ref{ec:dydlon}),
and~(\ref{ec:dxdlon}) we obtain
\begin{equation}
\frac{\partial y}{\partial\varphi}\frac{y}{s\,\pi}
=\frac{a^2(1-e^2)}{(1-e^2\sin^2\varphi)^2}\cos\varphi,
\end{equation}
which by separation of variables can be then integrated as
\begin{equation}
\label{ec:Integral}
\frac{1}{s\,\pi}\int y\,dy
=a^2\int\frac{(1-e^2)}{(1-e^2\sin^2\varphi)^2}\cos\varphi\,d\varphi.
\end{equation}
The primitive integral of the left side is straightforward, while for the right
side the solution can be seen in \cite[p. 107]{osborne2013}. Then,
\begin{equation}
\label{ec:SolEcDifElip}
\frac{1}{2}\frac{1}{s\,\pi}y^2=\frac{1}{2}a^2(1-e^2)
\left[\frac{\sin\varphi}{1-e^2\sin^2\varphi}
+\frac{\arctanh(e\sin\varphi)}{e}\right]+C,
\end{equation}
where $C$ is the integration constant. At the equator ($\varphi=0$) we have that
$y=0$, so, substituting these values in~(\ref{ec:SolEcDifElip}) we obtain
\begin{equation}
0=0+C\longrightarrow C=0,
\end{equation}
and then
\begin{equation}
\label{ec:SolEcDif}
\frac{1}{a^2s\,\pi}y^2=(1-e^2)\left[\frac{\sin\varphi}{1-e^2\sin^2\varphi}
+\frac{\arctanh(e\sin\varphi)}{e}\right].
\end{equation}
But right part of this equation takes part in the definition of authalic
latitude, which is \citep[p. 107]{osborne2013}
\begin{equation}
\label{ec:DefBeta}
\beta=\arcsin\frac{q}{q_P},
\end{equation}
where
\begin{equation}
\label{ec:Defq}
q=(1-e^2)\left[\frac{\sin\varphi}{1-e^2\sin^2\varphi}
+\frac{\arctanh(e\sin\varphi)}{e}\right],
\end{equation}
and
\begin{equation}
\label{ec:Defqp}
q_P=q(\varphi=\pi/2)=1+\frac{1-e^2}{e}\arctanh(e).
\end{equation}
Then, from~(\ref{ec:SolEcDif}), (\ref{ec:DefBeta}) and (\ref{ec:Defqp}) we get
\begin{equation}
\label{ec:SolEcDifBeta}
\frac{1}{a^2s\,\pi}y^2=q_P\sin\beta,
\end{equation}
and from~(\ref{ec:SolEcDifBeta}) and~(\ref{ec:xGeneral}) we arrive at
\begin{equation}
\label{ec:ElipDirecto}
\begin{dcases}
x=a\Delta\lambda\sqrt{\frac{q_P}{s\,\pi}\sin|\beta|},\\
y=\pm a\sqrt{s\,\pi\,q_P\sin|\beta|},
\end{dcases}
\end{equation}
where the $+/-$ signs in $y$ coordinate correspond to northern and southern
hemispheres respectively.

\subsection{Inverse problem}

Taking into account formulae~(\ref{ec:ElipDirecto}) it can be deduced for
the inverse problem that
\begin{equation}
\label{ec:ElipInverso}
\begin{dcases}
\beta=\pm\arcsin\left(\frac{y^2}{a^2s\,\pi\,q_P}\right)
\text{\hspace{3mm}and\hspace{3mm}}\beta\longrightarrow\varphi,\\
\lambda=\frac{x}{a}\sqrt{\frac{s\,\pi}{q_P\sin|\beta|}}+\lambda_0,
\end{dcases}
\end{equation}
where the $+$ sign must be used for $y\geq0$ and the $-$ sign when $y<0$.
Transformation from $\beta$ to $\varphi$ were performed traditionally via an
iterative algorithm or through a series expansion, as it can be seen in
\citet[p. 16]{snyder1987}. For precision up to 64 bits roundoff error (variables
of type \texttt{double} in programming languages) in $\varphi\rightarrow\beta$
and $\beta\rightarrow\varphi$ conversions, series expansions are available in
\cite{karney2024}.

\section{Formulation for the sphere}

\subsection{Forward problem}

\begin{figure*}[htb]
\centering
\includegraphics[width=0.99\textwidth]{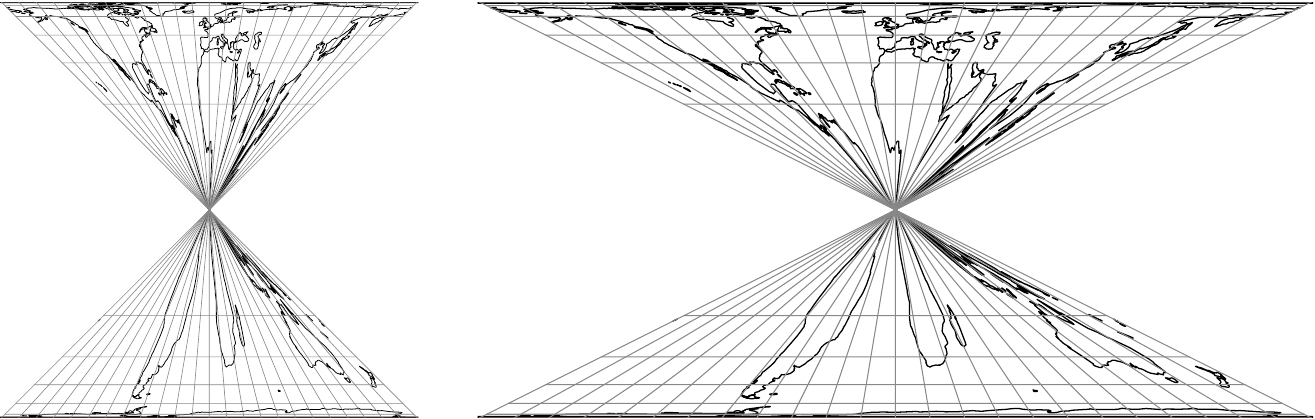}
\caption{Original \hourglass{} projection for the sphere with values $s=1$
         (left) and $s=0.5$ (right).}
\label{fig:figure3}
\end{figure*}

For the sphere $e=0$, so
\begin{equation}
q=\lim_{e\rightarrow0}(1-e^2)\left[\frac{\sin\varphi}{1-e^2\sin^2\varphi}
+\frac{\arctanh(e\sin\varphi)}{e}\right]=\sin\varphi+\frac{0}{0}.
\end{equation}
Then, applying the l'H\^opital rule,
\begin{equation}
\lim_{e\rightarrow0}\frac{\arctanh(e\sin\varphi)}{e}=\frac{0}{0}
=\lim_{e\rightarrow0}\frac{\frac{\sin\varphi}{1-e^2\sin^2\varphi}}{1}=\sin\varphi,
\end{equation}
so
\begin{equation}
\label{ec:LatAutEsfera}
\begin{dcases}
q=2\sin\varphi,\\
q_P=2
\end{dcases}
\longrightarrow\sin\beta=\frac{q}{q_P}=\sin\varphi\longrightarrow\beta=\varphi,
\end{equation}
and, from~(\ref{ec:ElipDirecto}),
\begin{equation}
\label{ec:DirectoEsfera}
\begin{dcases}
x=R\Delta\lambda\sqrt{\frac{2}{s\,\pi}\sin|\varphi|},\\
y=\pm R\sqrt{2s\,\pi\sin|\varphi|},
\end{dcases}
\end{equation}
where $R$ is the radius of the sphere, and match the form~(\ref{ec:Sny2})
and~(\ref{ec:Sny1}) as can be easily proved by derivation.
Figure~\ref{fig:figure3} shows the projection for the sphere and $s=1$ (left)
and $s=0.5$ (right).

\subsection{Inverse problem}

The inverse problem for the sphere is straightforward.
From~(\ref{ec:DirectoEsfera}) we obtain
\begin{equation}
\label{ec:InversoEsfera}
\begin{dcases}
\varphi=\pm\arcsin\left(\frac{y^2}{2R^2s\,\pi}\right),\\
\lambda=\frac{x}{R}\sqrt{\frac{s\,\pi}{2\sin|\varphi|}}+\lambda_0,
\end{dcases}
\end{equation}
where the $+$ sign must be used for $y\geq0$ and the $-$ sign when $y<0$.

\section{Generalized \hourglass{} projection}

\begin{figure}[htb]
\centering
\includegraphics[width=0.35\textwidth]{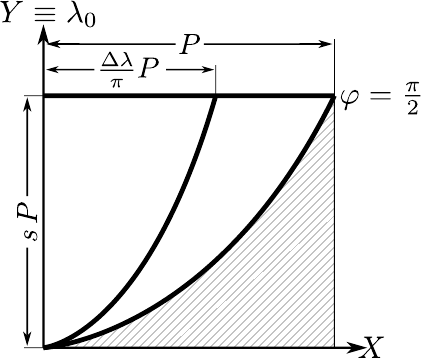}
\caption{Projection of the origin meridian $\lambda_0$ and the pole.}
\label{fig:figure4}
\end{figure}

Meridians in \hourglass{} projection as presented by John P. Snyder are straight
lines ($y=c\,x$) convergent in the equator. But we want to develop a generalization
of these lines in the form
\begin{equation}
\label{ec:ycxn}
y=c\,x^n,\text{ }\forall n\in\mathbb{R}^+-\{0\}.
\end{equation}
Figure~\ref{fig:figure4}, in a similar way as Figure~\ref{fig:figure2},
shows for the general version of the projection the origin meridian $\lambda_0$
and the length $P$ straight line representing the North pole. Again, we impose the 
$y$ coordinate of the pole as $y=s\,P$, so an aspect ratio of $s:1$ means that the
height of the whole map is $s$ times its whole width. For any point with
longitude $\Delta\lambda=\lambda-\lambda_0$ its $x$ coordinate at the line
representing the pole is $x=\Delta\lambda\,P/\pi$.

Parameter $c$ is in this case, based on Equation~(\ref{ec:ycxn}),
\begin{equation}
\label{ec:DefC}
c=\frac{y}{x^n}=\frac{\pi^ns\,P}{\Delta\lambda^nP^n}
=s\frac{\pi^n}{\Delta\lambda^n}P^{1-n},
\end{equation}
which is valid for any point in a particular meridian, i.~e., for any $\varphi$
and $\Delta\lambda=\text{const.}$ Then, putting together~(\ref{ec:ycxn})
and~(\ref{ec:DefC}) we can write
\begin{equation}
\label{ec:yGen}
y=s\frac{\pi^n}{\Delta\lambda^n}P^{1-n}x^n,
\end{equation}
and
\begin{equation}
\label{ec:xGen}
x=
\frac{\Delta\lambda}{\pi}\left(\frac{1}{s}P^{n-1}|y|\right)^{\frac{1}{n}},
\end{equation}
which for $n=1$ is~(\ref{ec:xGeneral}).

About the value of $P$ in Equations~(\ref{ec:yGen}) and~(\ref{ec:xGen}),
consider the computation of the area $A$ of the region limited by the $Y$ axis,
the line representing the pole and the meridian of $\Delta\lambda=\pi$ in
Figure~(\ref{fig:figure4}). Such area can be computed by substracting from the area
of the rectangle with sides $s\,P$ and $P$ the area $A_d$ of the dashed region in
Figure~(\ref{fig:figure4}), which is
\begin{equation}
A_d=s\,P^{1-n}\int_0^Sx^ndx=s\frac{P^{1-n}}{n+1}\left.x^{n+1}\right|_0^P
=s\frac{P^2}{n+1},
\end{equation}
so
\begin{equation}
\label{ec:AProy}
A=s\,P^2-s\frac{P^2}{n+1}=s\,P^2\frac{n}{n+1}.
\end{equation}

The \hourglass{} projection is equal-area, so Equation~(\ref{ec:AProy}) must be
equal to the area of a fourth of the terrestrial ellipsoid surface, that is
\citep[p. 107]{osborne2013}
\begin{equation}
A_{\text{e}/4}
=\frac{\pi}{2}a^2\left[1+\frac{1-e^2}{e}\arctanh(e)\right],
\end{equation}
where $a$ is the semi major axis, $e^2=f(2-f)$ the squared first eccentricity,
and $f$ the flattening. Provided that $A=A_{e/4}$ we have
\begin{equation}
P=a\sqrt{\frac{1}{s}\frac{\pi}{2}\frac{n+1}{n}
\left[1+\frac{1-e^2}{e}\arctanh(e)\right]}.
\end{equation}
For the sphere of radius $R$ a fourth of its surface is
$A_{\text{s}/4}=\pi\,R^2$, so
\begin{equation}
P=R\sqrt{\frac{\pi}{s}\frac{n+1}{n}}.
\end{equation}

If we introduce a new parameter called $L$ we can write
\begin{equation}
\begin{dcases}
P=\frac{a}{\sqrt{s}}L,\text{ with }L=\sqrt{\frac{\pi}{2}\frac{n+1}{n}
\left[1+\frac{1-e^2}{e}\arctanh(e)\right]}\text{ for the ellipsoid},\\
P=\frac{R}{\sqrt{s}}\,L,\text{ with }L=\sqrt{\pi\frac{n+1}{n}}\text{ for the sphere},
\end{dcases}
\end{equation}
so Equation~(\ref{ec:xGen}) can be written as
\begin{equation}
\label{ec:xGenDef}
\begin{dcases}
x=\frac{\Delta\lambda}{\pi}\left(a^{n-1}s^{-\frac{n+1}{2}}L^{n-1}|y|\right)^{\frac{1}{n}}
\text{ for the ellipsoid},\\
x=\frac{\Delta\lambda}{\pi}\left(R^{n-1}s^{-\frac{n+1}{2}}L^{n-1}|y|\right)^{\frac{1}{n}}
\text{ for the sphere},
\end{dcases}
\end{equation}
which for $n=1$ are~(\ref{ec:xGeneral}).

\subsection{Forward problem on the ellipsoid}

The equal-area condition~(\ref{ec:CondEquiv}) must be fulfilled for the
generalized projection. The projected parallels for the generalized projection
are straight lines parallel to the $X$ axis, so $\partial y/\partial\lambda=0$,
as in Equation~(\ref{ec:dydlon}). Concerning the partial derivative of $x$ with
recpect to $\lambda$ we can write from Equation~(\ref{ec:xGen})
\begin{equation}
\frac{\partial x}{\partial\lambda}
=\frac{1}{\pi}\left(a^{n-1}s^{-\frac{n+1}{2}}L^{n-1}y\right)^{\frac{1}{n}},
\end{equation}
so the diferential equation is
\begin{equation}
\frac{\partial y}{\partial\varphi}
\frac{1}{\pi}\left(a^{n-1}s^{-\frac{n+1}{2}}L^{n-1}y\right)^{\frac{1}{n}}
=\frac{a^2(1-e^2)}{(1-e^2\sin^2\varphi)^2}\cos\varphi,
\end{equation}
which by separation of variables can be then integrated as
\begin{equation}
\label{ec:IntegralGen}
\frac{1}{\pi}\left(a^{n-1}s^{-\frac{n+1}{2}}L^{n-1}\right)^{\frac{1}{n}}\int y^{\frac{1}{n}}\,dy
=a^2\int\frac{(1-e^2)}{(1-e^2\sin^2\varphi)^2}\cos\varphi\,d\varphi.
\end{equation}
The integral of the right side is the same as in Equation~(\ref{ec:Integral}).
For the left side we have
\begin{equation}
\label{ec:IntIzqGen}
\int y^{\frac{1}{n}}\,dy=\frac{n}{n+1}y^{\frac{n+1}{n}}+C.
\end{equation}

Then, considering the integral of the right side of~(\ref{ec:Integral}), and the
Equations~(\ref{ec:DefBeta}), (\ref{ec:Defq}), (\ref{ec:Defqp}), (\ref{ec:xGenDef}),
(\ref{ec:IntegralGen}) and~(\ref{ec:IntIzqGen}) we arrive to
\begin{equation}
\label{ec:ElipDirectoGen}
\begin{dcases}
x=\frac{a}{\sqrt{s}}\frac{\Delta\lambda}{\pi}\left(\frac{\pi}{2}\frac{n+1}{n}L^{n-1}q_P\sin|\beta|\right)^{\frac{1}{n+1}},\\
y=\pm a\sqrt{s}\left(\frac{\pi}{2}\frac{n+1}{n}L^{\frac{1-n}{n}}q_P\sin|\beta|\right)^{\frac{n}{n+1}},
\end{dcases}
\end{equation}
where the $+/-$ signs in $y$ coordinate correspond to northern and southern
hemispheres respectively. If we use $n=1$ we obtain
Equation~(\ref{ec:ElipDirecto}).

\subsection{Inverse problem on the ellipsoid}

Taking into account the formulation~(\ref{ec:ElipDirectoGen}) it can be deduced
that
\begin{equation}
\label{ec:ElipInversoGen}
\begin{dcases}
\beta=\pm\arcsin\left[\left(\frac{|y|}{a\sqrt{s}}\right)^{\frac{n+1}{n}}
\frac{2}{\pi}\frac{n}{n+1}L^{\frac{n-1}{n}}\frac{1}{q_P}\right]
\text{\hspace{3mm}and\hspace{3mm}}\beta\longrightarrow\varphi,\\
\lambda=\frac{x}{a}\pi\sqrt{s}\left(\frac{2}{\pi}\frac{n}{n+1}L^{1-n}
\frac{1}{q_P\sin|\beta|}\right)^{\frac{1}{n+1}}+\lambda_0,
\end{dcases}
\end{equation}
where the $+$ sign must be used for $y\geq0$ and the $-$ sign when $y<0$. If we
use $n=1$ we obtain Equation~(\ref{ec:ElipInverso}).

\subsection{Forward problem on the sphere}

\begin{figure*}[htb]
\centering
\includegraphics[width=0.99\textwidth]{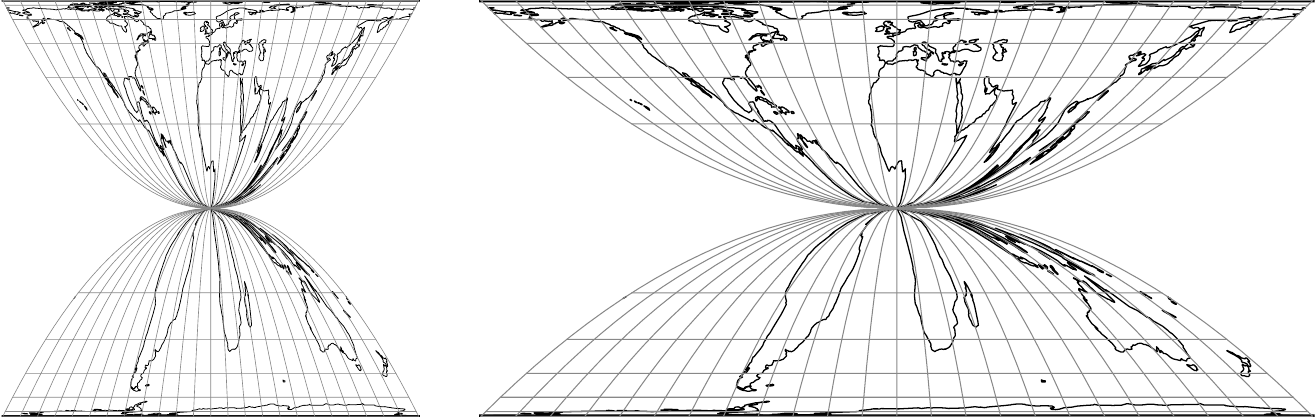}
\caption{Hourglass projection for the sphere with values $n=2$, $s=1$
         (left) and $s=0.5$ (right).}
\label{fig:figure5}
\end{figure*}

\begin{figure}[htb]
\centering
\includegraphics[width=0.95\textwidth]{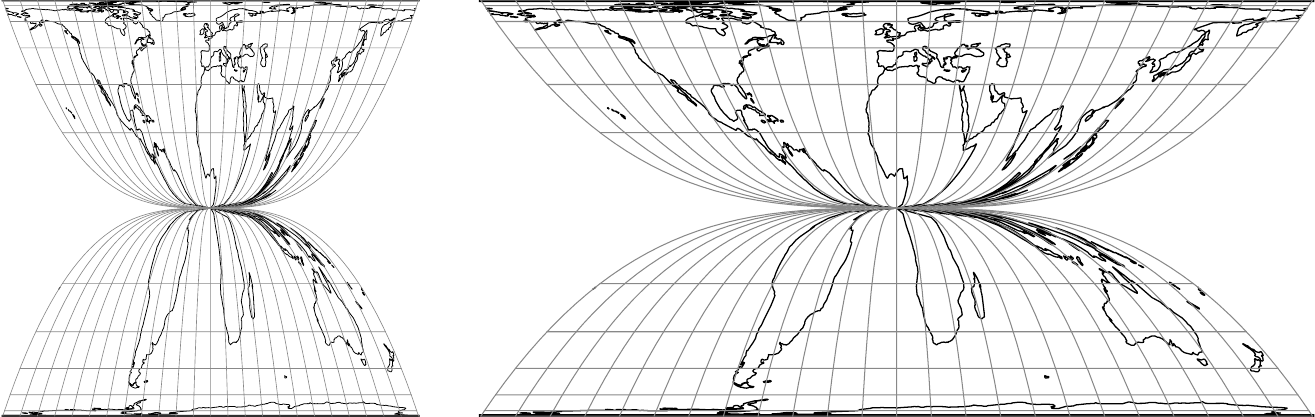}
\caption{Hourglass projection for the sphere with values $n=3$, $s=1$
         (left) and $s=0.5$ (right).}
\label{fig:figure6}
\end{figure}

For the sphere, taking into account Equation~(\ref{ec:LatAutEsfera}) and the
second expression in Equation~(\ref{ec:xGenDef}) the equal-area
condition~(\ref{ec:CondEquiv}) derives in
\begin{equation}
\label{ec:EsfDirectoGen}
\begin{dcases}
x=\frac{R}{\sqrt{s}}\frac{\Delta\lambda}{\pi}\left(\pi\frac{n+1}{n}L^{n-1}\sin|\varphi|\right)^{\frac{1}{n+1}},\\
y=\pm R\sqrt{s}\left(\pi\frac{n+1}{n}L^{\frac{1-n}{n}}\sin|\varphi|\right)^{\frac{n}{n+1}},
\end{dcases}
\end{equation}
where $R$ is the radius of the sphere and the $+/-$ signs in $y$ coordinate
correspond to northern and southern hemispheres respectively. If we use $n=1$ we
obtain Equation~(\ref{ec:DirectoEsfera}). Figure~\ref{fig:figure5} shows the
projection for the sphere, $n=2$ and $s=1$ (left) and $s=0.5$ (right).
Figure~\ref{fig:figure6} shows the projection for the sphere, $n=3$ and $s=1$
(left) and $s=0.5$ (right). Equations~(\ref{ec:EsfDirectoGen}) also have the
form~(\ref{ec:Sny2}) and~(\ref{ec:Sny1}) as can be proved by derivation.

\subsection{Inverse problem on the sphere}

Taking into account the formulation~(\ref{ec:EsfDirectoGen}) it can be deduced
that
\begin{equation}
\label{ec:EsfInversoGen}
\begin{dcases}
\varphi=\pm\arcsin\left[\left(\frac{|y|}{R\sqrt{s}}\right)^{\frac{n+1}{n}}
\frac{1}{\pi}\frac{n}{n+1}L^{\frac{n-1}{n}}\right],\\
\lambda=\frac{x}{R}\pi\sqrt{s}
\left(\frac{1}{\pi}\frac{n}{n+1}L^{1-n}\frac{1}{\sin|\varphi|}\right)^{\frac{1}{n+1}}+\lambda_0,
\end{dcases}
\end{equation}
where the $+$ sign must be used for $y\geq0$ and the $-$ sign when $y<0$. If we
use $n=1$ we obtain Equation~(\ref{ec:InversoEsfera}).

Note that for the inverse problem coordinate $\lambda$
in~(\ref{ec:EsfInversoGen}), as well as in  Equations~(\ref{ec:ElipInversoGen}),
(\ref{ec:InversoEsfera}), and~(\ref{ec:ElipInverso}), is undefined for points on
the equator (single point in the map):
\begin{equation}
\varphi=\beta=0\longrightarrow\sin\varphi=\sin\beta=0.
\end{equation}

\section{Calculation of standard parallels}

As it was stated in section~\ref{sec:Intro}, along a certain parallels $\pm\varphi_0$
distances in \hourglass{} projection are true. Then, the $x$ coordinate of a point
$Q=(\varphi_0,\pi)$ will be equal to half the perimeter of such parallel, so, for
the sphere and $n=1$ (Equation~(\ref{ec:DirectoEsfera})) we have
\begin{equation}
\label{ec:EcParaK}
R\,\pi\sqrt{\frac{2}{s\,\pi}\sin|\varphi_0|}=\pi\,R\cos\varphi_0\longrightarrow
\frac{2}{s\,\pi}\sin\varphi_0=\cos^2\varphi_0=1-\sin^2\varphi_0,
\end{equation}
whose solution is
\begin{equation}
\sin^2\varphi_0+\frac{2}{s\,\pi}\sin\varphi_0-1=0\longrightarrow
\sin\varphi_0=-\frac{1}{s\,\pi}\pm\sqrt{\frac{1}{s^2\pi^2}+1}
\end{equation}
where the only valid root is
\begin{equation}
\label{ec:Lat0Esf}
\sin\varphi_0=-\frac{1}{s\,\pi}+\sqrt{\frac{1}{s^2\pi^2}+1}.
\end{equation}
If we want to assign a specific standard parallel, the corresponding $s$
parameter can be computed from~(\ref{ec:EcParaK}) as
\begin{equation}
\label{ec:KEsf}
s=\frac{2}{\pi}\frac{\sin\varphi_0}{\cos^2\varphi_0}.
\end{equation}

Applying the same criterion for the ellipsoid (Equation~(\ref{ec:ElipDirecto}))
we have
\begin{equation}
\label{ec:EcParaFi0Elip}
a\,\pi
\sqrt{\frac{q_P}{s\,\pi}\sin|\beta_0|}=\pi\,\nu_0\cos\varphi_0\longrightarrow
s\,\pi\left(\frac{\nu_0}{a}\right)^2\cos^2\varphi_0-q_P\sin\beta_0=0,
\end{equation}
that can be solved by the Newton-Raphson method using as starting guess for
$\varphi_0$ the value obtained with~(\ref{ec:Lat0Esf}). If we want to assign a
specific standard parallel, the corresponding $s$ parameter can be computed from
Equation~(\ref{ec:EcParaFi0Elip}) as
\begin{equation}
\label{ec:KElip}
s=\left(\frac{a}{\nu_0}\right)^2\frac{q_P\sin\beta_0}{\pi\cos^2\varphi_0}.
\end{equation}
Table~\ref{tab:nkes} shows the different latitudes of standard parallels
corresponding to aspect ratios $2:1$, $1:1$ and $0.5:1$ on the ellipsoid (GRS80
parameters) and the sphere.

For the general case in the sphere the condition of $x$ coordinate
(Equation~(\ref{ec:EsfDirectoGen})) of a point $Q=(\varphi_0,\pi)$ be equal to
half the perimeter of $\varphi_0$ parallel leads to
\begin{equation}
\frac{R}{\sqrt{s}}\left(\pi\frac{n+1}{n}L^{n-1}\sin|\varphi_0|\right)^{\frac{1}{n+1}}
=\pi\,R\cos\varphi_0
\end{equation}
so
\begin{equation}
\frac{n+1}{n}\frac{L^{n-1}}{s^{\frac{n+1}{2}}\pi^n}\sin\varphi_0-\cos^{n+1}\varphi_0=0,
\end{equation}
which is Equation~(\ref{ec:EcParaK}) for $n=1$ and can be solved by the
Newton-Raphson method using as starting guess the value obtained
with~(\ref{ec:Lat0Esf}). The $s$ parameter corresponding to a specific latitude
$\varphi_0$ is
\begin{equation}
s=\left(\frac{L^{n-1}}{\pi^n}\frac{n+1}{n}
\frac{\sin|\varphi_0|}{\cos^{n+1}\varphi_0}\right)^{\frac{2}{n+1}},
\end{equation}
which is Equation~(\ref{ec:KEsf}) for $n=1$.

Applying the same criterion for the ellipsoid
(Equation~(\ref{ec:ElipDirectoGen})) we have
\begin{equation}
\frac{a}{\sqrt{s}}\left(\frac{\pi}{2}\frac{n+1}{n}L^{n-1}q_P\sin|\beta|\right)^{\frac{1}{n+1}}
=\pi\,\nu_0\cos\varphi_0,
\end{equation}
so
\begin{equation}
2\pi^n\left(\frac{\nu_0}{a}\sqrt{s}\right)^{n+1}\frac{n}{n+1}L^{1-n}\cos^{n+1}\varphi_0
-q_P\sin\beta_0=0,
\end{equation}
which is Equation~(\ref{ec:EcParaFi0Elip}) when $n=1$ and, as in precedent
cases, can be solved numerically using the Newton-Raphson method with the value
obtained in~(\ref{ec:Lat0Esf}) as starting guess. The $s$ parameter
corresponding to a specific latitude $\varphi_0$ is
\begin{equation}
\label{ec:sElip}
s=\left[\left(\frac{a}{\nu_0}\right)^{n+1}\frac{n+1}{n}
\frac{L^{n-1}}{2\pi^n}q_P
\frac{\sin|\beta_0|}{\cos^{n+1}\varphi_0}\right]^{\frac{2}{n+1}},
\end{equation}
which is Equation~(\ref{ec:KElip}) when $n=1$.

In practical implementation of all formulae for the ellipsoid presented in this
work is recommended to make the change
\begin{equation}
q_P\sin\beta=q,\text{\hspace{3mm}since\hspace{3mm}}
\sin\beta=\frac{q}{q_P}.
\end{equation}
Table~\ref{tab:nkes} shows the
different latitudes of standard parallels corresponding to $n=1$, $n=2$ and
$n=3$, and aspect ratios $2:1$, $1:1$ and $0.5:1$ on the ellipsoid (GRS80
parameters) and the sphere.
\begin{table}[htb]
\begin{center}
\caption{Latitude of standard parallels for different aspect ratios on the
         ellipsoid associated to the GRS80 reference system and the sphere.}
\begin{tabular}{c|c|c|c}
\label{tab:nkes}
$\mathbf{n}$ & $\mathbf{s:1}$ & \textbf{Ellipsoid} & \textbf{Sphere}\\
\hline
\multirow{3}{*}{$1$} & $2$   & $\pm\ang{58;42;35.0742}$ & $\pm\ang{58;35;12.5826}$\\\cline{2-4}
                     & $1$   & $\pm\ang{47;07;42.2058}$ & $\pm\ang{46;58;51.9387}$\\\cline{2-4}
                     & $0.5$ & $\pm\ang{33;26;09.6555}$ & $\pm\ang{33;17;11.7500}$\\\cline{2-4}
\hline
\multirow{3}{*}{$2$} & $2$   & $\pm\ang{62;09;12.9104}$ & $\pm\ang{62;02;35.7270}$\\\cline{2-4}
                     & $1$   & $\pm\ang{50;46;42.6119}$ & $\pm\ang{50;38;20.7398}$\\\cline{2-4}
                     & $0.5$ & $\pm\ang{35;35;01.6747}$ & $\pm\ang{35;25;51.4544}$\\\cline{2-4}
\hline
\multirow{3}{*}{$3$} & $2$   & $\pm\ang{63;29;35.4921}$ & $\pm\ang{63;23;17.8854}$\\\cline{2-4}
                     & $1$   & $\pm\ang{52;15;09.1489}$ & $\pm\ang{52;07;02.5328}$\\\cline{2-4}
                     & $0.5$ & $\pm\ang{36;18;24.7925}$ & $\pm\ang{36;09;11.0887}$\\
\end{tabular}
\end{center}
\end{table}

Finally, all formulation developed in this work has been implemented for PROJ
library \citep{evenden2025}, and could also be accessed from all software that
integrates PROJ, such as, e.~g., QGIS\footnote{\url{https://qgis.org}.} or
GDAL\footnote{\url{https://gdal.org}.}. Until its integration in PROJ's development 
master branch in the near future, it can be used from the branch ``hourglass'' from 
author \textsuperscript{a}.

\section{Practical evaluation}

Figures~\ref{fig:figure1}, \ref{fig:figure3}, \ref{fig:figure5}, 
and~\ref{fig:figure6} show, as it was stated, large distortions, specially near
the equator and the poles, making the \hourglass{} projection useless for world 
maps. However, near the standard parallels (northern and southern hemispheres) 
distortions can be tolerable in order to make useful maps, understanding these as 
those that maintain a good visual appearance that we try to assess. We consider as 
working area a tessera limited by parallels $\ang{45;;}\text{N}$ and 
$\ang{60;;}\text{N}$, and meridians $\ang{15}\text{W}$  and $\ang{15;;}\text{E}$.
Also, $\lambda_0=\ang{0;;}\text{E}$ is selected as origin meridian. We will check 
\hourglass{} projection for values $n=1,2,3$, and $s=1$ (see Table~\ref{tab:nkes} 
for the corresponfing standard parallels) in all cases.

We will compare \hourglass{} against two other equal-area 
projections: (1) Equal Earth \citep{savric2019} as an optimal pseudocylindrical world 
projection, and (2) the well-known Lambert Equal Area conic projection (LEAC) 
with $\varphi_0=\ang{52;30;}\text{N}$ as standard parallel, 
as representative of projection that is well adapted for zones \textit{wider} in 
longitude than \textit{taller} in latitude, as our test area is. 
Table~\ref{tab:resExample} shows the
results of the comparison, which was performed by computing the linear 
($k$) and angular ($\omega$) distortions in a point grid for all projections 
using PROJ software. As the projections are all equal-area, the linear and 
angular distortions depend on direction, so ``Avg.~$k$'' and ``Avg.~$\omega$'' 
fields in the table mean the average value of the maximum distortions for each 
point of the grid. In all cases a sphere of radius 
$R=\SI{6371000}{\metre}$ was used.
\begin{table}[htb]
\begin{center}
\caption{Comparison between Lambert Equal Area conic (LEAC), Equal Earth, and
         \hourglass{} projections in the defined working region and using the
         sphere of radius $R=\SI{6371000}{\metre}$.}
\begin{tabular}{c|c|c|c|c}
\label{tab:resExample}
\textbf{Projection} & \textbf{Avg.} $k$ & 
\textbf{Avg.} $\mathbf{\omega}$ ($^\circ$) & 
\textbf{Max.} $k$ & \textbf{Max.} $\mathbf{\omega}$ ($^\circ$)\\
\hline
\textbf{LEAC} & 1.0113 & 1.2823 & 1.02495 & 2.824 \\
\hline
\textbf{Equal Earth} & 1.1556 & 16.2649 & 1.30364 & 30.035 \\
\hline
\textbf{Hourglass} ($n=1$) & 1.1949 & 19.3813 & 1.48636 & 44.272 \\
\hline
\textbf{Hourglass} ($n=2$) & 1.1205 & 12.6679 & 1.31790 & 31.238 \\
\hline
\textbf{Hourglass} ($n=3$) & 1.1075 & 11.4592 & 1.25729 & 26.011 \\
\end{tabular}
\end{center}
\end{table}

\begin{figure}[htb]
\centering
\includegraphics[width=0.495\textwidth]{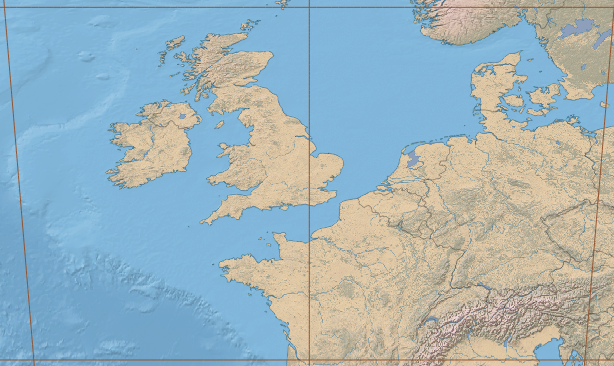}
\includegraphics[width=0.495\textwidth]{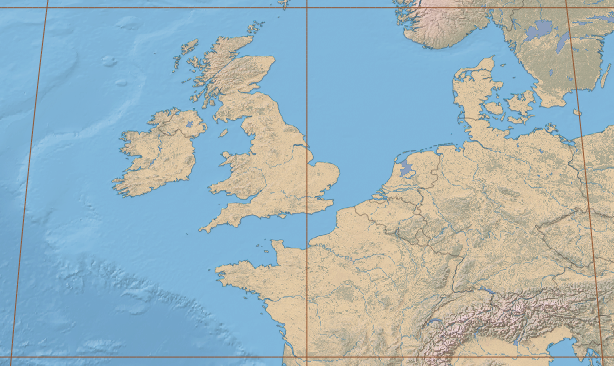}
\caption{Test area according to \hourglass{} projection with $n=1$ and $s=1$ (left),
         and according to Equal Earth projection (right).}
\label{fig:hourglass_eu}
\end{figure}

As it can be seen in results of Table~\ref{tab:resExample}, Lambert Equal Area conic
projection (LEAC) is the one that best suits the area under study, as it was expected 
given the shape of the test area. The Equal Earth projection shows linear and angular 
deformations greater than LEAC, which was expected as Equal Earth is optimal for the 
entire Earth as a whole, but not for an isolated portion. \hourglass{} projection shows
linear and angular deformations also greater than LEAC, but better than Equal Earth
depending on the parameter $n$. The original \hourglass{} proposed by Snyder ($n=1$, 
straigth projected meridians) presents greater deformations than Equal Earth, but when 
$n$ increases the deformations are lower than the ones of Equal Earth for our test area 
around the \hourglass' standard parallel. The change from $n=1$ to $n=2$ in \hourglass{}
results in a big decrease of deformations (around $\ang{7;;}$ in average angular 
deformation), which make the projection better suited for the test area than Equal 
Earth. When $n=3$ the deformations continue to decrease, but to a much lesser extent
than for the jump $n=1$ to $n=2$. Figure~\ref{fig:hourglass_eu} shows the working area
projected according \hourglass{} with $n=1$ (left) and according Equal Earth (right). As
it can be seen, shapes are well preserved in \hourglass{}, which make the map useful
when it is restricted to areas around the standard parallel. It can be also observed
the main feature that distinguishes \hourglass{} projection from other pseudocylindrical
projections: meridians in \hourglass{} converge to the Equator.


\section{Conclusions}

Direct and inverse formulations of the equal-area pseudocylindrical \hourglass{}
projection have been developed step by step from the mathematical point of view, and 
considering the ellipsoid and the sphere as reference surfaces. Moreover, the original 
projection proposed qualitatively by John P. Snyder in $1946$, i.e., $79$ years ago, 
has been generalized in order to allow meridians to follow any curve of the form 
$y=c\,x^n$  for $n\in\mathbb{R}^+-\{0\}$. Formulation has also been implemented for
PROJ package, a free software library and tools which is the de facto standard in map 
projections and datum transformations tasks.

Although the \hourglass{} projection was not proposed with the aim of using in real 
applications, and is not useful for world maps, we have proved that it could be 
applicable in mapping small areas around their standard parallels, providing linear
and angular deformations equal or lower than the ones for the Equal Earth projection.

\section*{Acknowledgements}

Special thanks to Iván Sánchez Ortega for revealing the existence of Snyder's Hourglass 
projection, and for challenging us to develop it in PROJ. Author \textsuperscript{b}
acknowledges the support of project \textit{Tecnolog\'ias en Ciencias del Patrimonio 
(TEC Heritage-CM, TEC-2024/TEC-39)} funded by the \textit{Tecnolog\'ias 2024} call by 
Comunidad Aut\'onoma de Madrid (Spain). Finally, we are truly grateful to all the people 
out there who make free software.

\bibliographystyle{tfx}
\bibliography{shaw2025}

\end{document}